\documentstyle[11pt]{article}
\begin{document}
\hoffset = -1truecm \voffset = -2truecm
\title{\bf Chaos of soliton systems \\
and special Lax pairs for chaos systems}
\author{Sen-yue Lou$^{1,2,3,4}$\thanks{Email: sylou@mail.sjtu.edu.cn},
Xiao-yan Tang$^{2,3}$ and Ying Zhang$^{3}$\\ \it \footnotesize \it
CCAST (World Laboratory), PO Box 8730, Beijing 100080, P. R.
China\\ \it \footnotesize \it Physics Department of Shanghai Jiao
Tong University, Shanghai 200030, P. R. China\thanks{Mailing
address}\\ \footnotesize \it $^{3}$Abdus Salam International
Centre for Theoretical Physics, Trieste, Italy}
\date{}

\maketitle

\begin{abstract}
In this letter, taking the well known (2+1)-dimensional soliton
systems,  Davey-Stewartson (DS) model and the asymmetric
Nizhnik-Novikov-Veselov (ANNV) model, as two special examples, we
show that some types of lower dimensional chaotic behaviors may be
found in higher dimensional soliton systems. Especially, we derive
the famous Lorenz system and its general form from the DS equation
and the ANNV equation. Some types of chaotic soliton solutions can
be obtained from analytic expression of higher dimensional soliton
systems and the numeric results of lower dimensional chaos
systems. On the other hand, by means of the Lax pairs of some
soliton systems, a \em lower \rm dimensional chaos system may have
some types of \em higher \rm dimensional Lax pairs. An explicit
(2+1)-dimensional Lax pair for a (1+1)-dimensional chaotic
equation is given.
\end{abstract}
\vskip.1in


\vskip.1in

In the past three decades, both the solitons ${\cite{soliton}}$
and the chaos${\cite{chaos}}$ have been widely studied and applied
in many natural sciences and especially in almost all the physics
branches such as the condense matter physics, field theory, fluid
dynamics, plasma physics and optics etc. Usually, one considers
that the solitons are the basic excitations of the integrable
models, and the chaos is the basic behavior of the nonintegrable
models. Actually, the above consideration may not be complete
especially in higher dimensions. When one says a model is
integrable, one should emphasize two important facts. The first
one is that we should point out the model is integrable under what
special meaning(s). For instance, we say a model is Painlev\'e
integrable if the model possesses the Painlev\'e property, and a
model is Lax or IST (inverse scattering transformation) integrable
if the model has a Lax pair and then can be solved by the IST
approach. An integrable model under some special meanings may not
be integrable under other meanings. For instance, some Lax
integrable models may not be Painlev\'e integrable$\cite{Eilbeck,
lsytxy}$. The second fact is that for the general solution of a
higher dimensional integrable model, say, a Painlev\'e integrable
model, there exist some lower dimensional \em arbitrary \rm
functions, which means any lower dimensional chaotic solutions can
be used to construct exact solutions of higher dimensional
integrable models.

In this letter, we will show that an IST integrable model and/or a
Painlev\'e integrable model may have some lower dimensional
reductions with chaotic behaviors and then a lower dimensional
chaos system may have some higher dimensional Lax pairs.

To show our conclusions, we use the (2+1)-dimensional
Davey-Stewartson (DS) equation${\cite{DS}}$
\begin{eqnarray}
&&i u_t+2^{-1}(u_{xx}+u_{yy})+\alpha |u|^2u-uv=0,\\ &&v_{xx}-
v_{yy}-2\alpha (|u|^2)_{xx}=0,
\end{eqnarray}
as a concrete example at first. The DS equation is an isotropic
Lax integrable extension of the well known (1+1)-dimensional
nonlinear Schr\"odinger (NLS) equation. The DS system is the
shallow water limit of the Benney-Roskes equation${\cite{DS}}$,
where $u$ is the amplitude of a surface wave-packet and $v$
characterizes the mean motion generated by this surface wave. The
DS system (1) and (2) can also be derived from the plasma
physics${\cite{NAS}}$ and the self-dual Yang-Mills field. The DS
system has also been proposed as a 2+1 dimensional model for
quantum field theory${\cite{SAB}}$. It is known that the DS
equation is integrable under some \em special \rm meanings,
namely, it is IST integrable and Painlev\'e
integrable$\cite{Boiti}$. Many other interesting properties of the
model like a special bilinear form, the Darboux transformation,
finite dimensional integrable reductions, infinitely many
symmetries and the rich soliton structures${\cite{Boiti, Lou}}$
have also been revealed.

To select out some chaotic behaviors of the DS equation, we make
the following transformation
\begin{eqnarray}
&&v =v_0-f^{-1}(f_{x'x'}+f_{y'y'}+2f_{x'y'})
+f^{-2}(f_{x'}^2+2f_{y'}f_{x'}+f_{y'}^2), \\ &&u = gf^{-1}+u_0
\end{eqnarray}
with real $f$ and complex $g$, where $x'=(x+y)/\sqrt{2},\
y'=(x-y)/\sqrt{2}$, and $\{u_0,\ v_0\}$ is an arbitrary seed
solution of the DS equation. Under the transformation (3) and (4),
the DS system (1) and (2) is transformed to a general bilinear
form:
\begin{eqnarray}
&&(D_{x'x'}+D_{y'y'}+2iD_t)g\cdot f
+u_0(D_{x'x'}+2D_{x'y'}+D_{y'y'})f\cdot f\nonumber\\ &&
\qquad+2\alpha u_0gg^* +2\alpha u_0^2g^*f-2v_0gf+G_1fg=0
\end{eqnarray}
\begin{eqnarray}
2(D_{x'y'}+\alpha |u_0|^2)f\cdot f +2\alpha gh+2\alpha g f
u_0^*+2\alpha u_0g^*f-G_1ff=0,
\end{eqnarray}
where $D$ is the usual bilinear operator${\cite{Hirota}}$ defined
as $D_{x}^mA\cdot B\equiv (\partial_x-\partial_{x_1})^m
A(x)B(x_1)|_{x_1=x}$, and $G_1$ is an arbitrary solution of
$-16\alpha(u_{0x'}+u_{0y'})(u_{0x'}^*+u_{0y'}^*)
+G_{1x'x'}+G_{1y'y'}+2G_{1x'y'}
-4\alpha(D_{x'x'}+D_{y'y'}+2D_{x'y'})u_0\cdot u_0^*=0.$ For the
notation simplicity, we will drop the ``primes" of the space
variables later.

To discuss further, we fix the seed solution $\{u_0,\ v_0\}$ and
$G_1$ as
\begin{eqnarray}
u_0=G_1=0, \qquad v_0=p_0(x,t)+q_0(y,t),
\end{eqnarray}
where $p_0\equiv p_0(x,t)$ and $q_0\equiv q_0(y,t)$ are some
functions of the indicated variables.

To solve the bilinear equations (5) and (6) with (7) we make the
ansatz
\begin{eqnarray}
f=C+p+q,\ g=p_1q_1\exp(ir+is),
\end{eqnarray}
where $p \equiv p(x,\ t) ,\ q\equiv q(y,\ t) ,\ p_1\equiv p_1(x,\
t) ,\ q_1\equiv q_1(y,\ t) ,\ r\equiv r(x,\ t),\ s\equiv s(y,\ t)
$ are all real functions of the indicated variables and $C$ is an
arbitrary constant. Substituting (8) into (5) and (6), and
separating the real and imaginary parts of the resulting equation,
we have
\begin{eqnarray}
2p_xq_y-\alpha p_1^2q_1^2=0.
\end{eqnarray}
\begin{eqnarray}
&&(q_1 p_{1xx} + p_1 q_{1yy} -p_1 q_1 (2r_t+2s_t +2 (p_0+q_0)
+s_{y}^2 +r_{x}^2 ))  (C+p+q) \nonumber\\ && + q_1 (p_1 p_{xx}-2
p_{1x} p_{x}) + p_1 (q_1 q_{yy}-2 q_{1y} q_{y})=0
\end{eqnarray}
\begin{eqnarray}
&&( -q_1 (2r_{x} p_{1x} +2 p_{1t} + p_1r_{xx}) - p_1 (2s_{y}
q_{1y} +2 q_{1t}+q_1 s_{yy}))(C+p+q) \nonumber\\ && +2 q_1 p_1
(q_t+s_{y} q_{y}) +2 q_1 p_1 (r_{x} p_{x}+p_t)=0
\end{eqnarray}

Because the functions $p_0,\ p,\ p_1$ and $r$ are only functions
of $\{x,\ t\}$ and the functions $q_0,\ q,\ q_1$ and $s$ are only
functions of $\{y,\ t\}$, the equation system (9), (10) and (11)
can be solved by the following variable separated equations:
\begin{eqnarray}
p_1=\delta_1\sqrt{2\alpha^{-1}c_1^{-1}p_x},\
q_1=\delta_2\sqrt{c_1q_y},\ (\delta_1^2=\delta_2^2=1),
\end{eqnarray}
\begin{eqnarray}
p_t=-r_{x}p_{x}+c_2,\ q_t=-s_{y}q_{y}-c_2,
\end{eqnarray}
\begin{eqnarray}
&& 4(2r_t+r_{x}^2+2p_0)p_{x}^2+p_{xx}^2
-2p_{xxx}p_{x}+c_0p_{x}^2=0,\\
&&4(2s_t+s_{y}^2+2q_0)q_{y}^2+q_{yy}^2
-2q_{y}q_{yyy}-c_0q_{y}^2=0.
\end{eqnarray}
In Eqs. (12)--(15), $c_1,\ c_2$ and $c_0$ are all arbitrary
functions of $t$.

Generally, for a given $p_0$ and $q_0$ the equation systems
\{(12), (14)\} and \{(13), (15)\} may not be integrable. However, because of
the arbitrariness of $p_0$ and $q_0$, we may treat the functions
$p$ and $q$ are arbitrary while $p_0$ and $q_0$ are determined by
(14) and (15). Because $p$ and $q$ are arbitrary functions, in
addition to the stable soliton selections, there may be various
chaotic selections. For instance, if we select $p$ and $q$ are
solutions of ($\tau_1\equiv x+\omega_1 t,\  \tau_2\equiv
x+\omega_2 t$)
\begin{eqnarray}
p_{\tau_1\tau_1\tau_1}=\left(p_{\tau_1\tau_1}p_{\tau_1}+(c+1)p_{\tau_1}^2\right)p^{-1}
-(p^2+bc+b)p_{\tau_1}-(b+c+1)p_{\tau_1\tau_1}+pc(ba-b-p^2),\
 \\
q_{\tau_2\tau_2\tau_2}=\left(q_{\tau_2\tau_2}q_{\tau_2}+(\gamma+1)q_{\tau_2}^2\right)q^{-1}
-(q^2+\beta\gamma+\beta)q_{\tau_2}-(\beta+\gamma+1)q_{\tau_2\tau_2}+qc(\beta\alpha-\beta-q^2),
\end{eqnarray}
where $\omega_1,\ \omega_2,\ a,\ b,\ c,\ \alpha,\ \beta$ and
$\gamma$ are all arbitrary constants, then
\begin{eqnarray}
&& c_0=c_2=0,\ r=-\omega_1 (x+\omega_1 t/2),\ s=-\omega_2
(y+\omega_2 t/2), \\
&&p_0=-4^{-1}\left({cp^3}{p_{\tau_1}^{-1}}+p^2-{bc}{p_{\tau_1}^{-1}}(a-1)p
+b(c+1)+(b+c+1){p_{\tau_1\tau_1}}{p_{\tau_1}^{-1}}\right.\nonumber\\
&&\qquad \left.+{p_{\tau_1\tau_1}^2}{2^{-1}p_{\tau_1}^{-2}}
-p^{-1}(p_{\tau_1}(c+1)+p_{\tau_1\tau_1})\right),\\
&&q_0=-4^{-1}\left({\gamma
q^3}{q_{\tau_2}^{-1}}+q^2-{\beta\gamma}{q_{\tau_2}}^{-1}(\alpha-1)q
+\beta(\gamma+1)+(\beta+\gamma+1){q_{\tau_2\tau_2}}{q_{\tau_2}^{-1}}\right.\nonumber\\
&&\qquad \left.+{q_{\tau_2\tau_2}^2}{2^{-1}q_{\tau_2}^{-2}}
-q^{-1}(q_{\tau_2}(\gamma+1)+q_{\tau_2\tau_2})\right).
\end{eqnarray}
Substituting (8) with (12)--(20) into (3) and (4), we get a
general solution of the DS equation
\begin{eqnarray}
&&u={\delta_1\delta_2\sqrt{2\alpha^{-1}p_{\tau_1}q_{\tau_2}}\exp(-i(\omega_1x
+\omega_2y+\frac12(\omega_1^2+\omega_2^2)t)} {(C+p+q)^{-1}},\\
&&v=p_0+q_0-{(q_{\tau_2\tau_2}
+p_{\tau_1\tau_1})}{(C+p+q)^{-1}}+{(q_{\tau_2}^2
+2q_{\tau_2}p_{\tau_1}+p_{\tau_1}^2)}{(C+p+q)^{-2}}
\end{eqnarray}
where $p_0$ and $q_0$ are determined by (19) and (20), while $p$
and $q$ are given by (16) and (17).

It is straightforward to prove that (16) (and (17)) is equivalent
to the well known chaos system, the Lorenz
system${\cite{Lorenz}}$:
\begin{eqnarray}
p_{\tau_1}=-c(p-g),\ g_{\tau_1}=p(a-h)-g,\ h_{\tau_1}=pg-bh.
\end{eqnarray}
Actually, after canceling the functions $g$ and $h$ in  (23), one
can find (16) immediately.

From the above discussions, some interesting things are worth emphasizing:

Firstly, because of the arbitrariness of the functions $p$ and
$q$, any types of other lower dimensional systems may be used to
construct the exact solutions of the DS system.

Secondly, the lower dimensional chaotic behaviors may be found in
many other higher dimensional soliton systems. For instance, by
means of the direct substitution or the similar discussions as for
the DS equation, one can find that ($\tau_1\equiv x+\omega_1t$)
\begin{eqnarray}
u={2p_{x}w_y}{(p+w)^{-2}},
\end{eqnarray}
\begin{eqnarray}
v=\frac{2p_{x}^2}{(p+w)^2}+\frac{(c+1)p_{x}}{3p}
-\frac{(5p-w)p_{xx}}{3p(p+w)}-\frac13 p^2
-\frac1{3p_{x}}[(1+b+c)p_{xx}+cp^3-cb(a-1)p],
\end{eqnarray}
with $w$ being an arbitrary function of $y$, and $p$ being
determined by the (1+1)-dimensional extension of the Lorenz system
\begin{eqnarray}
p_t=-p_{xxx}+p^{-1}[p_{xx}p_x+(c+1)p_x^2]-p^2p_x-(b+c+1)p_{xx}-pc(b-ba+p^2),
\end{eqnarray}
solves the following IST and Painlev\'e integrable KdV equation
which is known as the ANNV model${\cite{BLMP}}$
\begin{eqnarray}
u_t+u_{xxx}-3(uv)_x=0,\ u_x=v_y.
\end{eqnarray}
It is clear that the Lorenz system (16) is just a special
reduction of (27) with $ p=p(x+b(c+1)t)\equiv p(\tau_1). $
Actually, $p$ of equations (24) may also be arbitrary function of
$\{x,\ t\}$ after changing (25) appropriately $\cite{BLMP}$. In
other words, any lower dimensional chaotic behavior can also be
used to construct exact solutions of the ANNV system.

The third thing is more interesting.  The Lax pair plays a very
important and useful role in integrable models. Nevertheless,
there is little progress in the study of the possible Lax pairs
for chaos systems like the Lorenz system. In Ref.
$\cite{Eilbeck}$, Chandre and Eilbeck had given out the Lax pairs
for two discrete non-Painlev\'e integrable models. In Ref.
$\cite{lsytxy}$, we have found some Lax pairs for some
non-integrable Schwarzian equations. Now from the above
discussions, we know that both the Lax pairs of the DS equation
and those of the ANNV system may be used as the special higher
dimensional Lax pairs of \em arbitrary \rm chaos systems like the
Lorenz system (16) and/or the generalized Lorenz system (26) by
selecting the fields appropriately like (21)-(22) and/or
(24)-(25). For instance, the (1+1)-dimensional generalized Lorenz
system (26) has the following Lax pair
\begin{eqnarray}
\psi_{xy}=u\psi,\qquad \psi_t=-\psi_{xxx}+3v\psi_{x}
\end{eqnarray}
with $\{u,\ v\}$ being given by (24) and (25). From (24)-(27), we
know that a \em lower \rm dimensional chaos system can be
considered as a consistent condition of a \em higher \rm
dimensional linear system. For example, $\psi_{xyt}  =\psi_{txy} $
of (28) just gives out the generalized Lorenz system (26).

Now a very important question is what the effects of the lower
dimensional chaos to the higher dimensional soliton systems are.
To answer this question, we use the numerical solutions of the
Lorenz system to see the behaviors of the corresponding solution
(24) of the ANNV equation by taking $ w=200+\tanh(y-y_0)\equiv
w_s$ and $p=p(\tau_1)\equiv p(X)$ as the numerical solution of the
Lorenz system (16). Under the selection $w=w_s$, (24) is a line
soliton solution located at $y=y_0$. Due to the entrance of the
function $p$, the structures of the line soliton become very
complicated. For some types of the parameters, the solutions of
the Lorenz system have some kinds of periodic behavior, then the
line soliton solution (24) with $w=w_s$ becomes an $x$ periodic
line soliton solution that means the solution is localized in $y$
direction (not identically equal to zero only near $y=y_0$) and
periodic in $x$ direction. Fig. 1 shows the behavior of the
periodic two line soliton solution and $p$ being a periodic two
solution of the Lorenz system (16) with $a=350,\ b=8/3$ and
$c=10$. In some other types of the parameter ranges, the solution
of the Lorenz system becomes chaotic and then the line soliton of
the ANNV system becomes chaotic line soliton which is localized in
$y$ direction and chaotic in $x$ direction. Fig. 2 displays the
chaotic behavior of the amplitude of the line soliton located at
$y=y_0$ with $a=60, b=8/3$ and $c=10$. The parameters we used here
are the same as those used in literature$\cite{Lorenz}$.

\input epsf
\begin{figure}
\epsfxsize=7cm \epsfysize=6cm \epsfbox{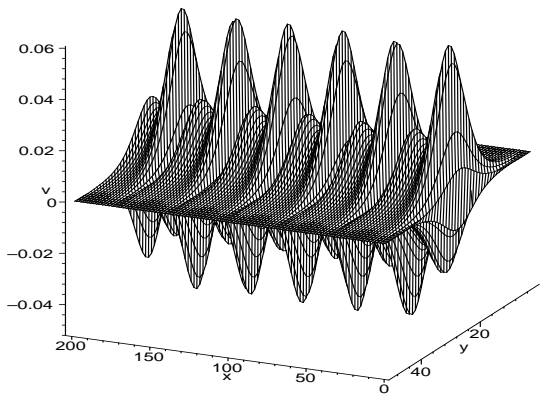}
\caption{Plot of the period two line soliton solution (24) of the
ANNV equation with (29), $p=p(\tau)$, and $a=350,\ b=8/3,\ c=10$
at $t=0$.}
\end{figure}
\input epsf
\begin{figure}
\epsfxsize=7cm \epsfysize=6cm \epsfbox{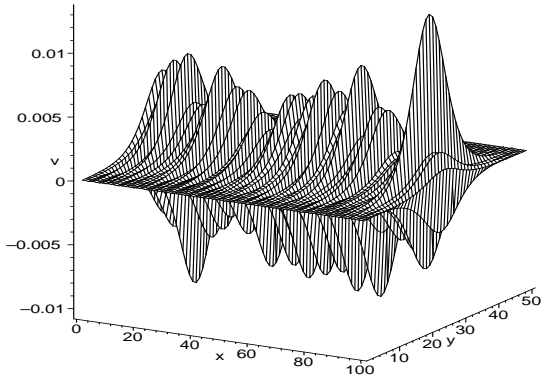}
\caption{Plot of the chaotic line soliton solution (24) of the
ANNV equation with the same conditions as Fig. 1 except for
$a=60$.}
\end{figure}

In summary, though some (2+1)-dimensional soliton systems, like
the DS equation and the ANNV equation, are Lax and IST integrable,
and some special types of soliton solutions can be found by IST
and other interesting approaches${\cite{Boiti}}$, any types of
chaotic behaviors may still be allowed in some special ways.
Especially, the famous chaotic Lorenz system and its
(1+1)-dimensional generalization are derived from the DS equation
and the ANNV equation. Using the numerical results of the lower
dimensional chaotic systems, we may obtain some types of nonlinear
excitations like the periodic and chaotic line solitons for higher
dimensional soliton systems. On the other hand, the lower
dimensional chaos systems like the generalized Lorenz system may
have some particular Lax pairs in higher dimensions.

It is also known that both the ANNV system and the DS systems are
related to the Kadomtsev-Petviashvili (KP) equation while the DS
and the KP equation are the reductions of the self-dual Yang-Mills
(SDYM) equation. So both the KP and the SDYM equations may possess
arbitrary lower dimensional chaotic behaviors induced by arbitrary
functions. From the results of this paper, various important and
interesting problems should be studied further. For instance, what
on earth is the \em complete \rm integrability and what kinds of
information about chaos can be obtained from some types of special
higher dimensional Lax pairs?

\vskip.2in The author is in debt to thanks the helpful discussions
with the professors Q. P. Liu and G-x Huang. The work was
supported by the National Outstanding Youth Foundation of China
(No.19925522), the Research Fund for the Doctoral Program of
Higher Education of China (Grant. No. 2000024832) and the National
Natural Science Foundation of Zhejiang Province of China.

\end{document}